\documentclass[prb,twocolumn,showpacs,superscriptaddress,floatfix, nofootinbib]{revtex4-2}

\usepackage{amsmath,amssymb,amsfonts,mathrsfs, amsbsy}

\usepackage{xcolor} 
\usepackage{graphicx}
\usepackage[bookmarks=true, colorlinks=true, linkcolor=blue, urlcolor=blue, citecolor=blue, bookmarks=true, hyperindex=true]{hyperref}
\usepackage[normalem]{ulem}
\usepackage{comment}
\usepackage{array}
\usepackage{float}  
\usepackage{makecell}
\usepackage{enumitem}

\definecolor{blueCustom}{HTML}{0072B2}   
\definecolor{orangeCustom}{HTML}{E69F00} 

\usepackage{tikz}

\newcommand{\be}{\begin{equation}}
\newcommand{\ee}{\end{equation}}

\newcommand{\beq}{\begin{eqnarray}}
\newcommand{\eeq}{\end{eqnarray}}

\def\H1{\widehat{H}_1}

\begin{document}

\title{Thermalization in classical systems with discrete phase space}



\newcommand{\ULFMF}{Department of Physics, Faculty of Mathematics and Physics, 
University of Ljubljana, Jadranska 21, SI-1000 Ljubljana, Slovenia}

\author{Pavel Orlov}
\affiliation{\ULFMF}

\author{Enej Ilievski}
\affiliation{\ULFMF}

\begin{abstract}
   We study the emergence of statistical mechanics in isolated classical systems with local interactions and discrete phase spaces. We establish that thermalization in such systems does not require global ergodicity; instead, it arises from effective local ergodicity, where dynamics in a subsystem may appear pseudorandom. To corroborate that, we analyze the spectrum of the unitary evolution operator and propose an ansatz to describe statistical properties of local observables expanded in the eigenfunction basis — the classical counterpart of the Eigenstate Thermalization Hypothesis. Our framework provides a unified perspective on thermalization in classical and quantum systems with discrete spectra.
\end{abstract}


\maketitle

\paragraph*{\textbf{Introduction}}---The main objective of statistical physics is to explain the emergence of statistical ensembles and macroscopic laws from the viewpoint of deterministic microscopic many-body dynamics. In spite of a recent progress in the domain of quantum systems, a theoretical framework for explaining the onset of thermalization in many-body systems that would conceptually unify classical and quantum dynamical systems has remained elusive so far.

Thermalization phenomena in classical systems are traditionally discussed through the prism of ergodic theory~\cite{Boltzmann1896,Birkhoff, cornfeld2012ergodic}. According to the ergodic hypothesis — a cornerstone of equilibrium statistical mechanics — a single trajectory eventually explores uniformly the accessible phase space,
ensuring that time averages along the trajectories coincide with ensemble averages. This perspective nevertheless suffers from several drawbacks \cite{Uffink}. On the one hand, explicit verification of ergodicity is a notoriously difficult computational problem for many realistic many-body systems, with rigorous results only available for certain special, idealized models \cite{Anosov1967,Sinai, bunimovich1979}. More prominently, ergodicity is a very stringent requirement: when obeyed, it implies thermalization for all observables, thus not explicitly distinguishing physically relevant, local, observables from highly nonlocal ones.

In the realm of quantum systems, thermalization is typically analyzed in terms of the spectral properties of the generator of the underlying unitary dynamics. This perspective constitutes the cornerstone of the celebrated Eigenstate Thermalization Hypothesis (ETH) \cite{rutkevich2012,Deutsch91,Srednicki94} which asserts that individual eigenstates of chaotic many-body Hamiltonians are indistinguishable from thermal states when probed by local observables.
Owing to ETH, the long-time averages of local observables match the predictions of statistical ensembles for a broad class of initial states. There is mounting evidence that the ETH holds in various quantum systems with discrete spectra, ranging from spin chains \cite{Rigol_2008,Steinigeweg,Steinigeweg_2014, RigolSrednicki,IkedaAllETH?,LevVidmar_2019, D_Alessio_2016} and quantum circuits~\cite{DeLuca,FelixProsen} to quantum field theories~\cite{Lashkari_2018,BasuThermalityCFT, ProsenSotiriadis}, cementing it as the central paradigm for explaining the onset of thermalization in quantum systems.

Classical dynamics, however, also admits a unitary formulation. In particular, the time evolution of $L^2$-integrable observables is generated by the unitary Koopman operator \cite{Koopman1931}, hinting that dynamical properties of classical and quantum systems may possibly admit a unified description. Unfortunately, this approach is obstructed in classical systems with continuous phase space, where the Koopman spectrum is generically continuous and largely inaccessible to direct analysis~\cite{Budi_Koopmanism,brunton2021}.

In this Letter, we sidestep this issue by focusing on classical many-body systems with a discrete phase space,
where the discrete spectrum of the Koopman operator can be fully described in terms of periodic orbits. Notably, ergodicity seldom holds
in such systems since the largest orbit typically occupies a finite (or even vanishing) fraction of the phase space. Yet, in local subsystems, different orbits may appear indistinguishable from random trajectories, leading to thermalization of local observables. To verify this scenario, we formulate a quantitative condition, analogous to the ETH ansatz in quantum systems, revealing a striking similarity between the microscopic mechanisms of thermalization in classical and quantum systems with discrete spectra.

\paragraph*{\textbf{Dynamics in discrete phase space}}---In this work, we consider classical one-dimensional lattice systems with discrete degrees of freedom evolving in discrete time.
Local degrees of freedom which can take $q$ values,  $x_{a}\in \mathbb{Z}_{q}= \{ 0,...,q-1 \}$, and thus the full configuration space is given by $X = \{ \boldsymbol{x} = (x_1, ..., x_L) \, | \, x_a \in \mathbb{Z}_q  \} \equiv \mathbb{Z}_{q}^{L}$, where $L$ is the system length.

The discrete-time dynamics of configurations is, then, governed by a map $F:X \rightarrow X$ which specifies the one-step update rule $\boldsymbol{x}_{t+1} = F(\boldsymbol{x}_t)$, $t \in \mathbb{Z}_{\geq0}$. We additionally require the dynamics to be invertible, implying that $F$ is a \emph{permutation} acting on $X$.

Classical observables belong to a finite-dimensional Hilbert space $\text{Fun}(X)$ of functions over the phase space $X$ of dimension $\text{dim} \,\text{Fun}(X) = q^{L}$, with a basis $\{\pi_{\boldsymbol{x}}\}_{\boldsymbol{x}\in X}$ acting on configurations as $\pi_{\boldsymbol{x}}(\boldsymbol{y}) = \delta_{\boldsymbol{x},\boldsymbol{y}}$.
In particular, any observable $\mathrm{A} \in \text{Fun}(X)$ can be represented as $\mathrm{A} = \sum_{\boldsymbol{x}} A_{\boldsymbol{x}} \pi_{\boldsymbol{x}}$, with $A_{\boldsymbol{x}}$ denoting $\mathrm{A}$ (note the font difference) evaluated in the configuration $\boldsymbol{x}$. We subsequently focus on physically relevant, local observables whose values depend only on the subconfiguration supported on a finite contiguous range of lattice sites.

The space of observables $\text{Fun}(X)$ is equipped with the inner product $(\mathrm{A}|\mathrm{B}) = \sum_{\boldsymbol{x}} A_{\boldsymbol{x}}^{*} B_{\boldsymbol{x}},$
which is the discrete analog of the $L^{2}$ inner product with the Liouville measure used in systems with continuous phase spaces. Accordingly, $A_{\boldsymbol{x}}=(\pi_{\boldsymbol{x}} | A) $.

The dynamical map $F$ may additionally possess local conservation laws, i.e. observables of the form $\mathrm{Q}=\sum_{a}\mathrm{q}_{a}$ obeying $Q_{\boldsymbol{x}} = Q_{F(\boldsymbol{x})}$, where local densities $\mathrm{q}_{a}$ are supported on a finite contiguous sublattice starting at position $a$~\cite{sublattice}. In the presence of a single local conservation law, the phase space $X$ decomposes into invariant isolevel sets $X_Q = \{ \boldsymbol{x} \in X \mid Q_{\boldsymbol{x}} = Q \}$ called charge sectors, $X = \bigcup_Q X_Q$. Generalization to multiple local conservation laws is straigthforward~\cite{quasilocal}.

\nocite{Ilievski_2016}
\nocite{sharipov2025}

The most general probability distribution in the phase space $\rho \in {\rm Fun}(X)$ can be written as $\rho = \sum_{\boldsymbol{x}} p_{\boldsymbol{x}} \pi_{\boldsymbol{x}}$, with probabilities $p_{\boldsymbol{x}}\geq 0$ obeying $\sum_{\boldsymbol{x}} p_{\boldsymbol{x}} = 1$. In the absence of local conservation laws, the long-time average of any \emph{local} observable is expected to thermalize to a microcanonical ensemble average, $A_{\rm mc} \equiv (\rho_{\rm mc}|\mathrm{A}) = \frac{1}{|X|} \sum_{\boldsymbol{x}} A_{\boldsymbol{x}}$, defined with respect to the uniform microcanonical measure $\rho_{\rm mc} \equiv (\sum_{{\bf x} \in X}\pi_{\bf x})/|X|$.
Since in the presence of a local conservation law $\mathrm{Q}$ each sector $X_Q$ is equipped with its own uniform measure,
$\rho_{\mathrm{mc}}^{(Q)} \equiv (\sum_{{\bf x}\in X_{Q}}\pi_{{\bf x}}) / |X_Q|$, thermalization phenomena can be accordingly analyzed in each charge sector independently. For brevity, we restrict our subsequent discussion to systems without any conservation laws. All the formulae can nonetheless be simply amended to the case of multiple conservation laws by an appropriate restriction to the charge sector.

\paragraph*{\textbf{Spectral decomposition of unitary dynamics}}---Time evolution of trajectories governed by $F$ carries over to observables, which can be compactly expressed in terms of the Koopman operator $\mathcal{U} : \text{Fun}(X) \rightarrow \text{Fun}(X)$,
\begin{equation}\label{FP-oper-classical}
    \mathcal{U}\mathrm{A} = \sum_{\boldsymbol{x}} A_{\boldsymbol{x}} \pi_{F^{-1}(\boldsymbol{x})} = \sum_{\boldsymbol{x}} A_{ F(\boldsymbol{x})} \pi_{\boldsymbol{x}},
\end{equation}
which is \emph{unitary} with respect to the inner product. Dynamics of $\mathrm{A}$ with respect to the initial probability distribution $\rho_0 = \sum_{\boldsymbol{x}} p_{\boldsymbol{x}}^{(0)} \pi_{\boldsymbol{x}}$ is thus encoded in the overlap 
\begin{equation}\label{dynamics-with-rho}
    A_{\rho_0} (t) = ( \rho_{0} | \mathcal{U}^{t} \mathrm{A}),
\end{equation}
enabling to study dynamical properties in terms of the spectrum of $\mathcal{U}$, as is customary in quantum models. Obtaining the eigenmode spectrum of $\mathcal{U}$ boils down to finding all \emph{periodic orbits} (i.e. cycles) of the permutation map $F$: the trajectory $\gamma = \{ \boldsymbol{x}_{j} \}_{j=1}^{T_{\gamma}}$ is a periodic orbit with the (fundamental) period $T_{\gamma}$ provided that $T_{\gamma}$ is the smallest integer time for which $F(\boldsymbol{x}_{T_{\gamma}}) = \boldsymbol{x}_1$.
Each orbit $\gamma$ then yields exactly $T_{\gamma}$ eigenfunctions of the Koopman operator,
\begin{equation}\label{Koopman-eigenfunctions}
    \phi_{k}^{(\gamma)} = \frac{1}{T_{\gamma}} \sum_{j=1}^{T_{\gamma}} e^{-i 2 \pi j k/T_{\gamma} } \pi_{\boldsymbol{x}_j}, 
\end{equation}
satisfying $\mathcal{U} \phi_{k}^{(\gamma)} = e^{-i \omega_{k}^{(\gamma)}} \phi_{k}^{(\gamma)}$, with  orbit frequencies $\omega_{k}^{(\gamma)} \equiv 2 \pi k / T_{\gamma} \in [0, 2\pi]$ for $k \in \{0,1,\ldots,T_{\gamma}-1\}$.
The eigenmode decomposition of $\mathcal{U}$ thus takes the form
\begin{equation}
    \mathcal{U} = \sum_{\gamma} \sum_{k=0}^{T_{\gamma}-1} e^{-i \omega_{k}^{(\gamma)}} T_{\gamma} \left|\phi_k^{(\gamma)}\right) \left(\phi_{k}^{(\gamma)}\right|.
\end{equation}
Note the extra factor of $T_{\gamma}$ which is due to normalization $(\phi_{k}^{(\gamma)}| \phi_{k}^{(\gamma)} ) = 1 / T_{\gamma}$. 

We proceed by casting the dynamics of $\mathrm{A}$ and the corresponding correlation functions in terms of the overlap (Fourier) coefficients
\begin{equation}\label{overlaps}
    A_{k}^{(\gamma)} = (\phi_{k}^{(\gamma)} | \mathrm{A}) = \frac{1}{T_{\gamma}} \sum_{j=1}^{T_{\gamma}} e^{i j \omega_{k}^{(\gamma)}} A_{\boldsymbol{x}_j},
\end{equation}
which, as we demonstrate later, play a role analogous to matrix elements in quantum systems. Using Eq.~\eqref{dynamics-with-rho}, the long-time average of observable $\mathrm{A}$ evolving from the initial state $\rho_{0}$, namely $\overline{A_{\rho_0}}\equiv \lim_{t \rightarrow \infty}\frac{1}{t} \sum_{t'=0}^{t}  A_{\rho_0}(t')$, lies inside the invariant subspace of $\mathcal{U}$,
\begin{equation}\label{long-time-aver}
    \overline{A_{\rho_0}} = \sum_{\gamma} T_{\gamma} (\rho_0| \phi_0^{(\gamma)} ) A_0^{(\gamma)}.
\end{equation}
Note that eigenmodes $\phi_{0}^{(\gamma)}$ represent uniform measures supported on $\gamma$-orbits, whereas the weighted overlaps $T_{\gamma} (\rho_0 | \phi_{0}^{(\gamma)} ) = \sum_{\boldsymbol{x} \in \gamma} p_{\boldsymbol{x}}^{(0)}$ coincide with the probabilities that the initial configuration belongs to orbit $\gamma$. Formula \eqref{long-time-aver} can indeed be thought of as the classical analogue of the diagonal ensemble in quantum systems \cite{Rigol_2008, Polkovnikov_2011}. Unlike in the quantum case, the physical interpretation is rather obvious here: the long-time average of a local observable $\mathrm{A}$ evolving from an initial state belonging to $\gamma$-orbit is given by the zero mode $A_0^{(\gamma)}$ -- the average value of $\mathrm{A}$ over all the configurations in the set $\gamma$. 

While $A_{k\neq 0}^{(\gamma)}$ are not relevant for determining the long-time averages, their values are nevertheless crucial for characterizing the approach to equilibrium. Introducing the orbit averaging,
\begin{equation}\label{averaging}
    \mathbb{E}_{\gamma}[(\bullet)^{(\gamma)}] \equiv \sum_{\gamma} \nu_{\gamma} (\bullet)^{(\gamma)},\qquad \nu_{\gamma} \equiv T_{\gamma}/|X|,
\end{equation}
where $\nu_{\gamma}$ correspond to \emph{orbit fractions},
the dynamical two-point correlation function,
\begin{equation}\label{cor-function}
    C_{\mathrm{A}}(t) = \frac{1}{|X|}(\mathrm{A}|\mathcal{U}^{t}\mathrm{A}) = \frac{1}{|X|} \sum_{\boldsymbol{x}} A_{\boldsymbol{x}}(t) A_{\boldsymbol{x}}(0),
\end{equation}
can be expressed as $C_{\mathrm{A}}(t) = \mathbb{E}_{\gamma}[C_{\mathrm{A}}^{(\gamma)}(t)]$. Here $C^{(\gamma)}_{\mathrm{A}}(t) = \frac{1}{T_{\gamma}} \sum_{\boldsymbol{x} \in \gamma} A_{\boldsymbol{x}}(t) A_{\boldsymbol{x}}(0)$ denotes the correlation function restricted to orbit $\gamma$, whose power-spectrum is directly related to $|A_{k}^{(\gamma)}|^2$ via
\begin{equation}\label{orbit-corr-function}
\begin{aligned}
    C_{\mathrm{A}}^{(\gamma)}(t)  = \sum_{k=0}^{T_{\gamma}-1} |A_k^{(\gamma)}|^2 e^{-i \omega_{k}^{(\gamma)} t}.
\end{aligned}
\end{equation}

\paragraph*{\textbf{Orbit thermalization}}---The time-averaged observables $\overline{A_{\rho_{0}}}$, see Eq.~\eqref{long-time-aver}, still explicitly depend on microscopic details through the initial condition $\rho_{0}$. This information may however be irretrievably lost upon restricting to local observables acting within a subsystem $\Lambda$. To examine this scenarion and explain how $\overline{A_{\rho_0}}$ can be reconciled with the statistical ensemble prediction, we formulate an ansatz, Eqs.~\eqref{random-orbit-ETH} and \eqref{F-function}, which constitutes a \emph{classical} counterpart of the quantum ETH.
As a natural first step, we study an ensemble of random orbits.

\paragraph*{Random orbits.}
Since values of observables supported on $\Lambda$ depend only on the corresponding subconfigurations,
we consider an ensemble of random orbits of a given period $T_{\gamma}$ by sampling unbiased Bernoulli sequences $\{ \boldsymbol{x}_{j}^{(\Lambda) } \}_{j=1}^{T_\gamma}$ from $\mathbb{Z}_{q}^{|\Lambda|}$.
The central limit theorem ensures that, for large periods $T_{\gamma}$, the overlaps in Eq.~\eqref{overlaps} take the asymptotic form
\begin{equation}\label{random-orbit-ETH}
    A_{k}^{(\gamma)} = A_{\textrm{mc}} \delta_{k,0} + T^{-1/2}_{\gamma}R_k^{(\gamma)},
\end{equation}
where $R_k^{(\gamma)} \in \mathbb{C}$ (and $R_0^{(\gamma)} \in \mathbb{R}$ ) are \emph{random iid} Gaussian variables with mean $\mathbb{E}[R_{k}^{(\gamma)}] = 0$ and variance $\sigma_{A}^{2}\equiv \mathbb{E}[|R_k^{(\gamma)}|^{2}] = (A^2)_{\rm mc} - A_{\rm mc}^2$, respectively.

Equation \eqref{random-orbit-ETH} is formally analogous to the structure of matrix elements in the random basis in the quantum case, upon
identifying the Hilbert space dimension with the orbit size $T_{\gamma}$.

Although such algebraic proximity of $A^{(\gamma)}_{0}$ to the microcanonical value $A_{\rm mc}$ ensures thermalization, the flat power-spectrum implies white-noise correlations in Eq.~\eqref{orbit-corr-function}, that is $ C_{\rm A}(t ) = A_{\rm mc}^2 + \sigma_{A}^2 \delta_{t,0}$, $ 0\leq t \leq T_{\gamma}-1$.

\paragraph*{Deterministic orbits.}
Although individual trajectories in realistic systems governed by local deterministic evolution laws cannot be truly random, one can nonetheless expect them to effectively behave as random when probed by local measurements, and accordingly displaying statistical properties analogous to $A_k^{(\gamma)}$ in the random-orbit model, Eq.~\eqref{random-orbit-ETH}.

Typically, the mean orbit length
$T \equiv \mathbb{E}_{\gamma}[T_\gamma]$ in realistic discrete systems exhibits exponential growth with the system size, $ T \sim e^{\alpha L}$,  see, e.g., \cite{sharipov2025}. Then, assuming the scaling
$ A_0^{(\gamma)} - A_{\mathrm{mc}} \sim T_\gamma^{-1/2}$ holds, the orbit averages approach the microcanonical values exponentially fast with $L$ and, consequently, the long-time averages \eqref{long-time-aver} agree with the microcanonical prediction for a broad class of initial states.

One possibility to characterize the proximity of $A_{0}^{(\gamma)}$ to $A_{\mathrm{mc}}$ quantitatively is to analyze the scaling of the mean deviation $\text{MD}[\mathrm{A}] = \mathbb{E}_{\gamma}[ |A_{0}^{(\gamma)} - A_{\textrm{mc}}|]$ with the system size $L$ or the mean orbit length $T$. While a closely related quantity is used in the studies of diagonal ETH in quantum systems~\cite{Beugeling_2014,Steinigeweg_2014}, an important difference is that the average~\eqref{averaging}
accounts for finite orbit fractions; by contrast, the uniform average in quantum systems is taken over eigenstates that correspond to projectors of rank one.

Another utility of $\text{MD}[\mathrm{A}]$ is to bound the probability of deviations from $A_{\rm mc}$ under a random sampling of the initial condition ${\boldsymbol{x}}_{0}$ via $ \mathbb{P}(|\overline{A_{\boldsymbol{x}_0}} - A_{\textrm{mc}}| \geq \varepsilon) \leq \text{MD}[\mathrm{A}]/\varepsilon$.

In distinction to random orbits, however, nontrivial dynamical correlations will now be reflected in the statistical properties of \emph{pseudorandom} variables $R_{k}^{(\gamma)}$. Upon the frequency-window averaging
$\mathbb{E}_{\omega, \delta \omega}$ over all $k$ for which $\omega_{k}^{(\gamma)} \in [\omega, \omega + \delta \omega]$, the moments of $R^{(\gamma)}_{k}$ are expected to exhibit nontrivial $\omega$-dependence in the thermodynamic limit.
In particular, provided that the second moment
\begin{equation}\label{F-function}
   \lim_{\delta \omega \rightarrow 0} \lim_{L\rightarrow\infty}  \mathbb{E}_{\gamma} \mathbb{E}_{\omega, \delta \omega} [ |R_{k}^{(\gamma)}|^2 ] = F_{\rm A}(\omega),
\end{equation}
exist, it encodes the spectral weight of the thermodynamic correlation function $C_{\rm A}^{\rm th}(t)=\lim_{L\to \infty}C_{\rm A}(t)$ (cf. Eq.~\eqref{cor-function}):
\begin{equation}
    C_{\rm A}^{\textrm{th}}(t) = A_{\text{mc}}^2 + \frac{1}{2\pi} \int_{0}^{2\pi} d \omega e^{-i \omega t} F_{\rm A}(\omega).
\end{equation}

\paragraph*{Subsystem perspective.}
An alternative approach to study thermalization of local subsystems, which bypasses verifying the ansatz \eqref{random-orbit-ETH} for individual observables, is to instead directly analyze the structure of eigenfunctions \eqref{Koopman-eigenfunctions}, analogously to~\cite{Dymarsky_subsystem}. This approach invokes the projection operator $\mathcal{P}_{\Lambda}: \text{Fun}(X) \rightarrow \text{Fun}(X_{\Lambda})$ which reduces an observable ${\rm B}\in \text{Fun}(X)$ to a sublattice $\Lambda$ by `integrating out' (or marginalizing when acting on probability distributions) the complement $\overline{\Lambda}$ using the prescription
$(\mathcal{P}_{\Lambda} \mathrm{B})_{\boldsymbol{x}_{\Lambda}} \equiv \sum_{ \boldsymbol{x}_{\overline{\Lambda}} } B_{\boldsymbol{x}_{\Lambda}\cup \boldsymbol{x}_{\overline{\Lambda}}}$, akin to partial tracing in quantum mechanics. Therefore, for a local observable ${\rm A}$ supported on $\Lambda$, one can compute the overlaps using the reduced eigenfunctions, $A_{k}^{(\gamma)} = (\mathcal{P}_{\Lambda}\phi_{k}^{(\gamma)}|\mathrm{A})$.
The deviation of $A_0^{(\gamma)}$ from the microcanonical value can then be conveniently bounded by the $1$-norm distance,
\begin{equation}\label{deviation-bound}
    |A_{0}^{(\gamma)} - A_{\rm mc} | \leq \text{max}_{\boldsymbol{x}}|A_{\boldsymbol{x}}| \cdot d^{(\gamma)}_{\Lambda},
\end{equation}
where $d^{(\gamma)}_{\Lambda} \equiv || \mathcal{P}_{\Lambda}( \phi_{0}^{(\gamma)} - \rho_{\rm mc})||$ with $||\mathrm{B}|| = \sum_{\boldsymbol{x}} |B_{\boldsymbol{x}}|$. This reduces the analysis to the study of the \emph{average mean distance}
\begin{equation}\label{mean-distance}
    d_{\Lambda} \equiv \mathbb{E}_{\gamma} [d_{\Lambda}^{(\gamma)}],
\end{equation}
which quantifies the proximity of orbit distributions to the microcanonical ensemble within a subsystem. For instance, for an ensemble of random orbits with period $T$ one obtains the asymptotic law $d_{\Lambda} = \sqrt{\tfrac{2}{\pi}(q^{|\Lambda|}-1)}T^{-1/2}$.

The overlaps $A_{k\neq 0}^{(\gamma)}$ can likewise be upper-bounded as
\begin{equation}
    |A_{k}^{(\gamma)}| \leq \text{max}_{\boldsymbol{x}}|A_{\boldsymbol{x}}| \cdot || \mathcal{P}_{\Lambda} \phi_{k}^{(\gamma)} ||,
\end{equation}
which, assuming the asymptotic scaling $||\mathcal{P}_{\Lambda} \phi_{k}^{(\gamma)}|| \sim T_{\gamma}^{-1/2}$, ensures the validity of Eq.~\eqref{random-orbit-ETH} with $k\neq0$ for any observable supported on $\Lambda$. To verify this scaling numerically, we inspect whether the function
\begin{equation}\label{G-function}
    G_{\Lambda}(\omega; \delta \omega, L) \equiv \mathbb{E}_{\gamma} \mathbb{E}_{\omega,\delta \omega}[ T_{\gamma}^{1/2} || \mathcal{P}_{\Lambda} \phi_{k}^{(\gamma)} || ],
\end{equation}
admits a well-defined thermodynamic limit, $G_{\Lambda}(\omega)=\lim_{\delta \omega\to 0}\lim_{L\to \infty}G_{\Lambda}(\omega;\delta \omega,L)$. While for an ensemble of random orbits the $G$-function is just an $\omega$-independent constant $G_{\Lambda} (\omega) = \sqrt{ \pi q^{|\Lambda|}/2 }$, in realistic systems it can non-trivially depend on $\omega$, similarly to Eq.~\eqref{F-function}.

\paragraph*{\textbf{Numerical analysis}}---Although decomposing the dynamics into all periodic orbits requires computational resources that scale exponentially in system size, periodic orbits can nevertheless be efficiently sampled: by drawing an initial configuration $\boldsymbol{x}_0$ uniformly at random (thus selecting orbit $\gamma$ with probability $\nu_\gamma$) one simply traces the trajectory until its closes. Repeating this procedure yields an efficient approximation to $\mathbb{E}_{\gamma}$, see Eq.\eqref{averaging}.

To numerically test our predictions we subsequently specialize to a particular class of one-dimensional circuit models built from a local two-body update rule $f:\mathbb{Z}_q^2 \rightarrow \mathbb{Z}_q^2$.
Assuming $L$ is even, and adopting the periodic boundary conditions,
the global map $F$ is composed from two layers, $F = F_e \circ F_o$, given by the composition of $f$-maps, $F_o = f_{1,2} \circ f_{3,4} \circ \cdots \circ f_{L-1,L},$ and $F_e = f_{2,3} \circ f_{4,5} \circ \cdots \circ f_{L,1},
$ respectively (here $f_{i,j}$ denotes the action of $f$ on sites $(i,j)$ and identity elsewhere). Similar "brickwork" circuits are commonly employed in the study of discrete-time quantum dynamics and Floquet systems, see e.g. Refs.~\cite{Fisher_2023,bertini2025review}.

\textit{Models}. We pick two representative models with $q=3$, referred to as \texttt{Model I} and \texttt{Model II}, respectively, with the following update rules~\cite{sharipov2025,kim2025}:

\begin{tikzpicture}[
    box/.style={
        draw,
        minimum width=1.4cm,
        minimum height=1.1cm,
        align=center,
        font=\footnotesize
    }
]

\node (a00) at (0,0)  {\footnotesize $00 \to 00$};
\node (a01) at (1.3,0) {\footnotesize  $01 \to 01$};
\node (a02) at (2.6,0) {\footnotesize  $02 \to 21$};

\node (a10) at (0,-0.4) {\footnotesize $10 \to 22$};
\node (a11) at (1.3,-0.4) {\footnotesize $11 \to 02$};
\node (a12) at (2.6,-0.4) {\footnotesize $12 \to 11$};

\node (a20) at (0,-0.8) {\footnotesize $20 \to 20$};
\node (a21) at (1.3,-0.8) {\footnotesize $21 \to 10$};
\node (a22) at (2.6,-0.8) {\footnotesize $22 \to 12$};

\node (b00) at (4.3,0)  {\footnotesize $00 \to 00$};
\node (b01) at (5.6,0) {\footnotesize $01 \to 01$};
\node (b02) at (6.9,0) {\footnotesize $02 \to 10$};

\node (b10) at (4.3,-0.4) {\footnotesize $10 \to 02$};
\node (b11) at (5.6,-0.4) {\footnotesize $11 \to 12$};
\node (b12) at (6.9,-0.4) {\footnotesize $12 \to 22$};

\node (b20) at (4.3,-0.8) {\footnotesize $20 \to 20$};
\node (b21) at (5.6,-0.8) {\footnotesize $21 \to 11$};
\node (b22) at (6.9,-0.8) {\footnotesize $22 \to 21$};

\draw[draw=blueCustom, rounded corners=1pt, line width=1pt] (-0.6,0.2) rectangle  (0.6+2.6, -0.8 - 0.2) ;

\draw[draw=orange!65,rounded corners=1pt, line width=1pt] (4.3-0.6,0.2) rectangle  (4.3+0.6+2.6, -0.8 - 0.2) ;

\node at (5.6,0.38) { \texttt{Model II}};
\node at (1.3,0.38) { \texttt{Model I} };

\end{tikzpicture}

In \texttt{Model I}, as a consequence of time-reversal symmetry, the mean orbit length scales asymptotically as $T = \mathbb{E}_{\gamma}[T_{\gamma}] \sim \sqrt{|X|} =q^{L/2}$ (already indicating the lack of ergodicity in the usual sense), see Ref.~\cite{sharipov2025}.

\begin{figure}[h]
\centering
\includegraphics[width=0.9\columnwidth]{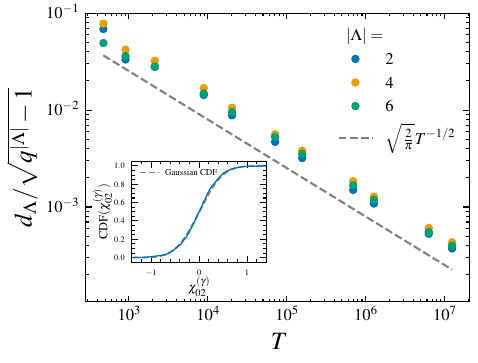}
\caption{Scaling of mean distance $d_{\Lambda}$ (rescaled by $\sqrt{q^{|\Lambda|}-1}$) with the mean orbit length $T$ in \texttt{Model I}, shown for various subsystem sizes. The largest $T$ corresponds to $L=28$. Inset: cumulative probability density function of fluctuations in the configuration frequencies, Eq.~\eqref{chi}, compared to a Gaussian fit.}
\label{fig:ModelI-distance-period}
\end{figure}

In \texttt{Model I} the mean distance scales asymptotically as $d_{\Lambda}\sim q^{|\Lambda|/2} T^{-1/2} $ (or $d_{\Lambda} \sim q^{|\Lambda|/2} q^{-L/4}$, see Fig.~\ref{fig:ModelI-distance-period}), confirming our hypothesis. 
Additionally, we studied fluctuations in the frequency $p_{\boldsymbol{s}} = T_{\gamma}^{-1} \sum_{j=1}^{T_{\gamma}} \boldsymbol{1}_{\{\boldsymbol{x}^{(\Lambda)}_{j} = \boldsymbol{s}\}} $ of a particular configuration $\boldsymbol{s} \in \mathbb{Z}_{q}^{|\Lambda|}$ defined as
\begin{equation}\label{chi}
    \chi_{\boldsymbol{s}}^{(\gamma)} = T^{1/2}_{\gamma} (p_{\boldsymbol{s}}^{(\gamma)} - q^{-|\Lambda|}).
\end{equation}
In the random-orbit ensemble, this quantity follows a Gaussian distribution.
As shown in the inset of Fig.~\ref{fig:ModelI-distance-period}, this also holds in \texttt{Model I}.

\begin{figure}[h]
\centering
\includegraphics[width=0.9\columnwidth]{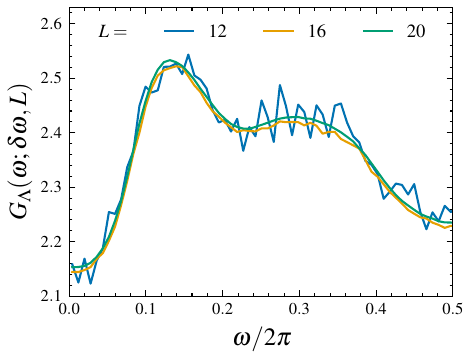}
\caption{The finite-size $G$-function, cf. Eq.\eqref{G-function}, for \texttt{Model I} for different system sizes $L$, shown for a subsystem of size $|\Lambda|=2$ and frequency window $\delta \omega = 0.05$.}
\label{fig:ModelI-Gfunction}
\end{figure}

Figure \ref{fig:ModelI-Gfunction} illustrates how the finite-size $G$-function \eqref{G-function} tends towards a smooth limiting function in the large-$L$ limit; its non-trivial frequency dependence indicates systematic deviations from the random-orbit ensemble.

\begin{figure}[h]
\centering
\includegraphics[width=0.9\columnwidth]{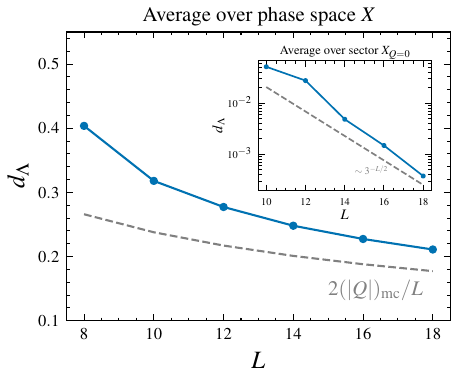}
\caption{Mean distance $d_{\Lambda}$ as a function of the system size $L$ in \texttt{Model II}, shown for a subsystem of size $|\Lambda| =2$.}
\label{fig:ModelII}
\end{figure}

In \texttt{Model II}, the mean orbit length scales as $T \sim |X| = q^{L}$, whereas the mean distance shows a slower, algebraic decay with $L$, see Fig.~\ref{fig:ModelII}. 

The algebraic decay is attributed to the fact that \texttt{Model II} possesses one local conservation law $\mathrm{Q} = \sum_{a=1}^{L/2} \mathrm{q}_{2a-1,2a}$, with local density $\mathrm{q}_{2a-1,2a}$ (acting as a number of $1$s on site $2a-1$ minus number of $0$s on site $2a$). Applying Eq.~\eqref{deviation-bound} to the local density (obeying $\langle \mathrm{q}_{2a-1,2a} \rangle_{\textrm{mc}} =0$ and $\max_{\boldsymbol{x}} |\mathrm{q}_{2a-1,2a}| =1$), and using the translational symmetry, it is easy to show that the mean distance can be bounded as
\begin{equation}\label{bound}
    d_{\Lambda} \geq \frac{ 2  (|\mathrm{Q}|)_{\textrm{mc}} } { L  } \approx \frac{4}{3 \sqrt{\pi}} \frac{1}{L^{1/2}}.
\end{equation}
Restricting to a charge sector $X_{Q}$ and replacing $\rho_{\rm mc}$ with $\rho_{\mathrm{mc}}^{(Q)}$, on the other hand, recovers the exponential decay $d_{\Lambda} \sim T^{-1/2} \sim q^{-L/2}$ (inset in Fig.~\ref{fig:ModelII}).

We emphasize that in the presence of a conserved local charge, the vanishing of $d_{\Lambda}$ from the uniform ensemble $\rho_{\mathrm{mc}}$ is a merely a corollary of the random sampling of initial conditions: with probability approaching $1$, such sampling selects a macrostate of maximal entropy corresponding to the zero density of $Q$; but the corresponding microcanonical ensemble is locally indistinguishable from $\rho_{\mathrm{mc}}$. Meanwhile, the fact that fluctuations of $Q$, owing to its locality, scale as $L^{1/2}$, yields the bound Eq.~\eqref{bound}. Similar algebraic decay has been observed in the studies of ETH even in integrable quantum systems \cite{Ikeda_2013_weakintegrableETH,Alba_2015_weakintegrableETH}.

Such a `measure concentration' mechanism is known in the literature under the name `typicality of thermalization'~\cite{Goldstein_2006,Nandy_2016,Cattaneo_2025, Cocciaglia_2022}. The latter, however, has nothing to do with the core mechanism of thermalization itself, which was the main subject of our work.

\paragraph*{\textbf{Conclusion}}---By investigating the conditions for thermalization in classical systems with discrete phase spaces, we established that emergence of statistical mechanics does not require global ergodicity, but instead relies on the weaker, effective local ergodicity within finite subsystems where dynamics can look pseudorandom.

We have shown how pseudorandomness can be diagnosed through the spectral properties of the unitary evolution operator by proposing an ansatz
for the expansion coefficients of local observables in the eigenfunction basis -- the classical counterpart of the Eigenstate Thermalization Hypothesis. Our formalism thus provides a unified approach to thermalization in classical and quantum systems with discrete spectra. The key differences between the two settings are largely attributed to the distinction between commuting and noncommuting algebras of observables.

There are several open problems and aspects that stand pending.
Since many recent studies of quantum ETH emphasize the importance of `asymptotic freeness'~\cite{Pappalardi_2022,Alves_2025,Pappalardi_2025,fritzsch2025}, it would be insightful to explore asymptotic independence and classical cumulants within our framework. Another important direction is to extend our analysis to integrable models featuring an extensive number of local charges, which have received a great deal of attention in the quantum domain~\cite{LeBlond_2019,essler2023,rottoli2025,OngoingWork}.

Finally, we wish to underline the fact that the present spectral approach to thermalization is only applicable to systems with discrete spectra.
Obtaining a quantitative framework which would also encompass classical or quantum systems with continuous spectra still remains a major open challenge.

\paragraph*{Acknowledgements.}

We thank P. Claeys, F. Fritzsch, D. Horváth, and L. Zadnik for valuable comments on the manuscript. P.O. gratefully acknowledges M. Moriniere for constant support during the work on this project. This work was supported by the Research Program P1-0402 and Project N1-0368 funded by the Slovenian Research Agency (ARIS).

\bibliography{references}

\end{document}